\documentclass[final]{aipproc}
\usepackage{graphicx}

\layoutstyle{6x9}

\begin{document}
\title{Modeling the Local Warm/Hot Bubble}


\author{Dieter Breitschwerdt}
{address={Institut f\"ur Astronomie, University of Vienna, T\"urkenschanzstra{\ss}e 17,
A-1180 Vienna, Austria}}
\author{Miguel A. de Avillez}
{address={Department of Mathematics, University of \'Evora, R. Rom\~ao
Ramalho 59, 7000 \'Evora, Portugal}}
\author{Verena Baumgartner}
{address={Institut f\"ur Astronomie, University of Vienna, T\"urkenschanzstra{\ss}e 17,
A-1180 Vienna, Austria}}

\begin{abstract}
In this paper we review the modeling of the Local Bubble (LB) with special emphasis on the
progress we have made since the last major conference ``The Local Bubble and Beyond (I)''
held in Garching in 1997. Since then new insight was gained into the possible origin of the
LB, with a moving group crossing its volume during the last 10 - 15 Myr being most likely
responsible for creating a local cavity filled with hot recombining gas. Numerical high resolution 3D
simulations of a supernova driven inhomogeneous interstellar medium show that we can reproduce
both the extension of the LB and the O{\sc vi} column density in absorption measured with FUSE
for a LB age of 13.5 - 14.5 Myr. We further demonstrate that the LB evolves like an ordinary superbubble
expanding into a density stratified medium by comparing analytical 2D Kompaneets solutions to
Na{\sc i} contours, representing the extension of the local cavity. These results suggest that LB
blow-out into the Milky Way halo has occurred roughly 5 Myr ago.

\end{abstract}

\maketitle

\section{Introduction}
At the time of the last Local Bubble (LB) conference, ``The Local Bubble and Beyond (I)'', as it should
be called now, held in Garching in 1997, modeling of the LB was no more than guess work, because
the crucial boundary conditions, like e.g. how many supernovae exploded and when, were unknown (cf. \cite{cs01}).
Therefore the only constraint was to match EUV and soft X-ray data. In the meantime significant observational progress has been made, warranting detailed 3D high resolution numerical simulations of the origin and
evolution of the LB, which will be discussed below.

In an attempt to outline the morphology of the local cavity deficient of H{\sc i}, Na{\sc i} was used as a sensitive tracer in absorption against background stars \citep{sl99,lw03}. It has been found that the cavity is elongated and inclined by $\sim 20^\circ$ with respect to the Galactic midplane (similar to the Gould's Belt), and open towards the poles, suggesting it to be a so-called Local Chimney. These chimneys have been proposed long ago to be the result of spatially and temporally correlated supernova (SN) explosions \citep{ni89}, generating high pressure bubbles, which blow out of the Galactic disk. Meridional sections through the LB also show a variation of the contours with longitude, arguing for an inhomogeneous ambient medium. Since the extension of the soft X-ray emitting LB is much less certain known than that of the local cavity, we will for convenience not distinguish between the two in the following.

To establish the temperature and density of the LB from soft X-ray and EUV observations has been extremely difficult, because there was a noticeable deficit of lines at low energies, putting the standard assumption of collisional ionization equilibrium (CIE) into question. Subsequently, models based on non-equilibrium ionization (NEI) have been suggested by Breitschwerdt \& Schmutzler \citep{bs94}, which could explain the lack of lines by an overionized plasma due to delayed recombination of a rapidly adiabatically cooling superbubble, but were also proceeding from unknown initial conditions. While CIE models do not depend on the evolution and hence the thermodynamic path the plasma has taken, NEI models are crucially determined by the underlying astrophysical model. Hence, if a CIE model for known abundances, temperature and absorbing column density does not fit, CIE can be ruled out. Not so for NEI models. The failure of a particular NEI model just rules out that particular astrophysical model which was responsible for it, but not NEI altogether. However, it would be an exhausting task to devise series of NEI models trying to fit the data, without even knowing the initial conditions of the problem. The situation has been exacerbated by an as yet uncertain amount of soft X-rays produced locally by solar wind charge exchange reactions (SWCX) onto heliospheric neutrals (\cite{cr00}; see also Koutroumpa, Shelton, these proceedings). Although some protagonists of the SWCX contribution to soft X-rays seem to believe that nearly all of the emission in the Galactic plane could be accounted for, there are still photons from higher Galactic latitudes, which need to be explained, as well as the state of the then not X-ray emitting plasma, which has to fill the local cavity. In summary, it seems reasonable to assume that multiple SN explosions are responsible for a significant amount of the soft X-ray and UV emitting plasma, and to search for the origin of the ``smoking gun'', because there is no young stellar cluster found inside the LB. Investigating moving groups of the local ISM (Bergh\"ofer \& Breitschwerdt \cite{bb02}), which was later extended to a complete and unbiased analysis of Hipparcos data of a larger volume of 400 pc in diameter, and combining these data by ARIVEL catalogue data (Fuchs et al. \cite{fu06}), it turned out that indeed a moving group crossed the present day LB region about 10 - 15 Myr ago, with its surviving members being now part of the Sco Cen association (see also \cite{ma01}). During this passage about 14 to 19 SNe exploded and energized the LB. Although the trajectory of the moving group passes close to the LB boundary in the Galactic Center direction, it is not unrealistic that the expansion of the LB is predominantly in the anticenter direction, because the presence of the Loop I bubble, which has been generated almost coevally with the LB, strongly inhibits the LB expansion in the center direction. This effect has been observed by an annular X-ray shadow in the soft ROSAT bands due to a compressed interaction shell between the two bubbles with a column density of $\sim 7 \times 10^{20} \, {\rm cm}^{-2}$ \citep{ea95}.

Bearing these findings in mind, we were able to start high resolution 3D numerical simulations with well constrained initial conditions \citep{ba06}, and compare our results to recent observations, e.g. in the UV with FUSE. This review is organized as follows. In Section~2 we outline the physical and numerical setup of our LB model and discuss the results in Section 3. In Section 4, we show that basic features of the LB, like its extension, shell velocity etc. can also be derived from analytical 2D Kompaneets \citep{ko60} solutions. In Section 5 our summary and conclusions on LB modeling will be presented.

\section{Numerical Modeling}

Since stellar absorption line studies show variations in the shape of the local cavity, it is clear that the ambient medium into which the LB expands is inhomogeneous. We have therefore tried to set up a ``realistic'' environment from large scale SN driven ISM simulations (Avillez \& Breitschwerdt \cite{ab04}, \cite{ab05a}), which we evolved for about 200 Myr on a large Cartesian grid ($0\leq (x,y)\leq 1$ kpc size in the Galactic plane, and $-10\leq z \leq 10$ kpc perpendicular to it), sufficiently long to wipe out memory effects of initial conditions in the ISM setup, which are necessarily artificial. The LB was generated by successive explosions of massive stars from the Pleiades subgroup B1 ($\sim 19$ stars between 20 and 11.5 M$_{\odot}$), almost coevally with the Loop I superbubble, which is energized by SNe of the Sco Cen association ($\sim 39$ stars with masses between 31 and 14 M$_{\odot}$; cf. \cite{egg98}). Their times of explosion are calculated by a formula for the main sequence life time \citep{st72}. The trajectory of the stars has been calculated backwards in time using Hipparcos data for positions and proper motions, and ARIVEL data for radial velocities (for details see Bergh\"ofer \& Breitschwerdt \cite{bb02} and Fuchs et al. \cite{fu06}). The results are based on the solution of the hydrodynamic fluid equations, including: (i) a gravitational field of the stellar disk \citep{kg89}, (ii) radiative cooling assuming CIE and solar abundances (for $T \geq 10^{4}$ K we use Sutherland \& Dopita \citep{sd93} and for $10 < T < 10^{4}$ K Dalgarno \& McCray \citep{dm72}, corrected for solar abundances taken from Anders \& Grevesse \citep{ag89}), and (iii) uniform background heating due to starlight varying with $z$ \citep{wm95}. The LB runs were started after global dynamical equilibrium of the background ISM simulation was established (typically after 200 Myr of evolution time). The LB simulation time is then an additional 30 Myr, which comfortably exceeds the stellar cluster ages. For further details of the model setup, see Breitschwerdt \& Avillez \cite{ba06}.

\section{Results}
A prime result of all our simulations, both general and local ISM, is the key role of turbulence in governing the structure and non-linear evolution of a compressible high-Reynolds number medium. It affects the pressure and density distributions, the large amount of gas in intermediate temperature regimes (i.e.\ between ``phases'') etc., and should be kept in mind when interpreting the following results. It also looks as if a substantial amount of shear is generated by colliding gas streams, as well as by the break-out of gas from superbubbles.

The locally enhanced SN rate produced by the passage of the moving group leads initially to a coherent LB structure. As the bubble grows, its shape becomes irregular, because it feels the ambient density variations of an evolving background medium due to ongoing star formation, and it also develops internal structure at $t_{\rm evol} > 8$ Myr, where $t_{\rm evol}$ is the elapsed time after the first explosion. Both the LB and the co-evolving Loop~I superbubble are bounded by shells, which collide at about $t_{\rm evol}=10$ Myr. As a consequence, Rayleigh-Taylor instabilities (RTIs) develop (cf.\ Breitschwerdt et al. \cite{bef00}), which will lead to a complete fragmentation of the shell in about $3$ Myr from now. The present age of the LB can be inferred from a comparison to O{\sc vi} absorption line data (see below) and yields $t_{LB}=13.5 - 14.5$ Myr (see Fig.~\ref{fig1} (left)). With the last SN having exploded about 0.5 Myr ago, no more SNe are expected to occur in the LB, i.e. it is becoming extinct, and in about $10$ Myr the LB will be swallowed by Loop~I (see Fig.~\ref{fig1} (right)), which is still active at present.
\begin{figure}[ht]
\begin{minipage}[b]{0.5\linewidth}
\centering
\includegraphics[width=\hsize,clip=]{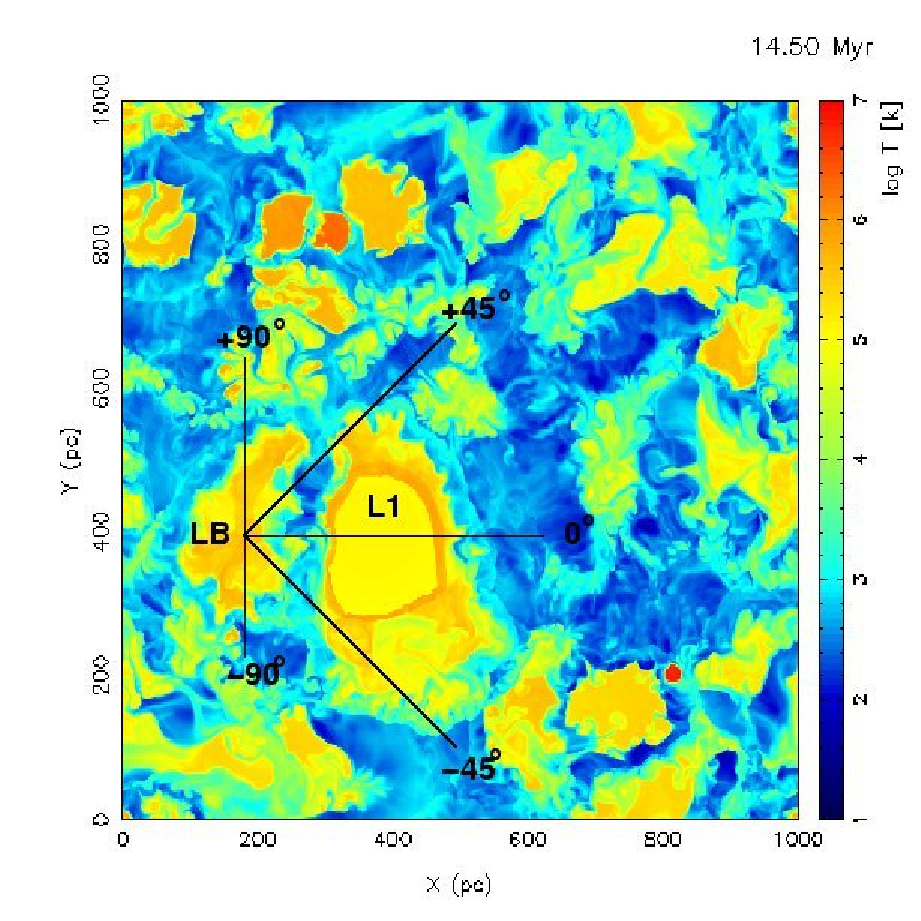}
\caption{default}
\end{minipage}
\hspace{0.5cm}
\begin{minipage}[b]{0.5\linewidth}
\centering
\includegraphics[width=\hsize,clip=,]{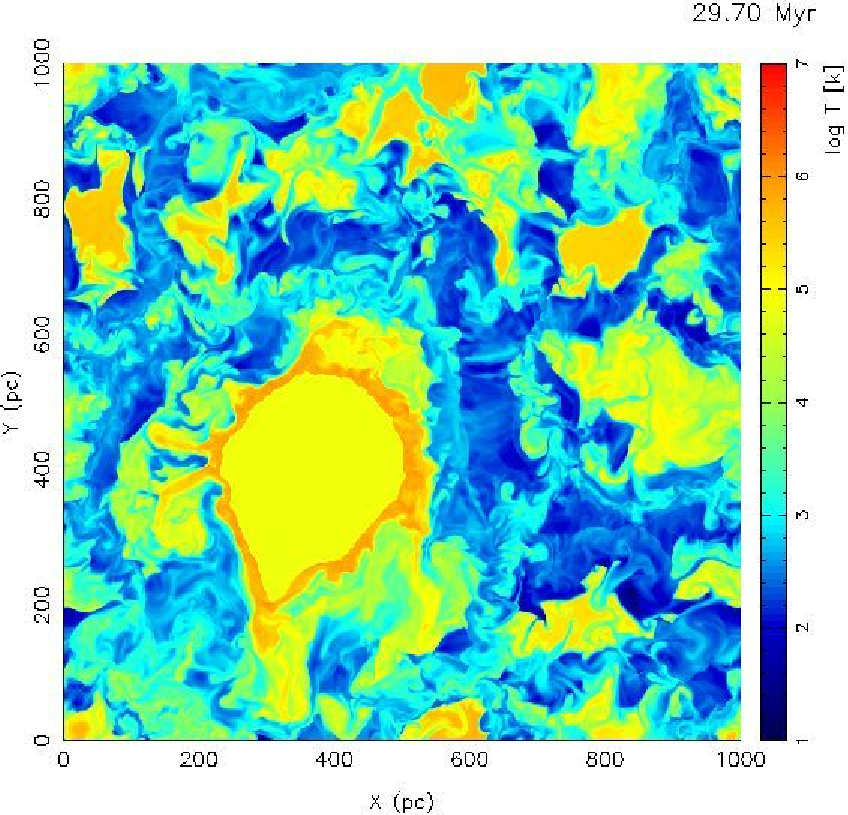}
\caption{default}
\end{minipage}
\caption{%
\textit{Left:} Color coded temperature maps in the
range $10 \leq T \leq 10^7$ K (taken from \cite{ba06}) for a slice through the data cube
(representing the Galactic midplane) of a 3D LB high resolution
simulation representing the present time (i.e.\ $\sim 14.5$ Myr after the
first explosion) with the LB centered at (175, 400) pc and Loop~I (L1) shifted
200 pc to the right. \textit{Right:} Same, showing the ``future'' of
the LB and L1 at $t=29.7$ Myr. Note that part of the LB has merged
with the ISM, part has been engulfed by L1.}
\label{fig1}
\end{figure}
%
%

\begin{figure}[htb]
  \centering
\includegraphics[width=0.7\hsize,angle=0]{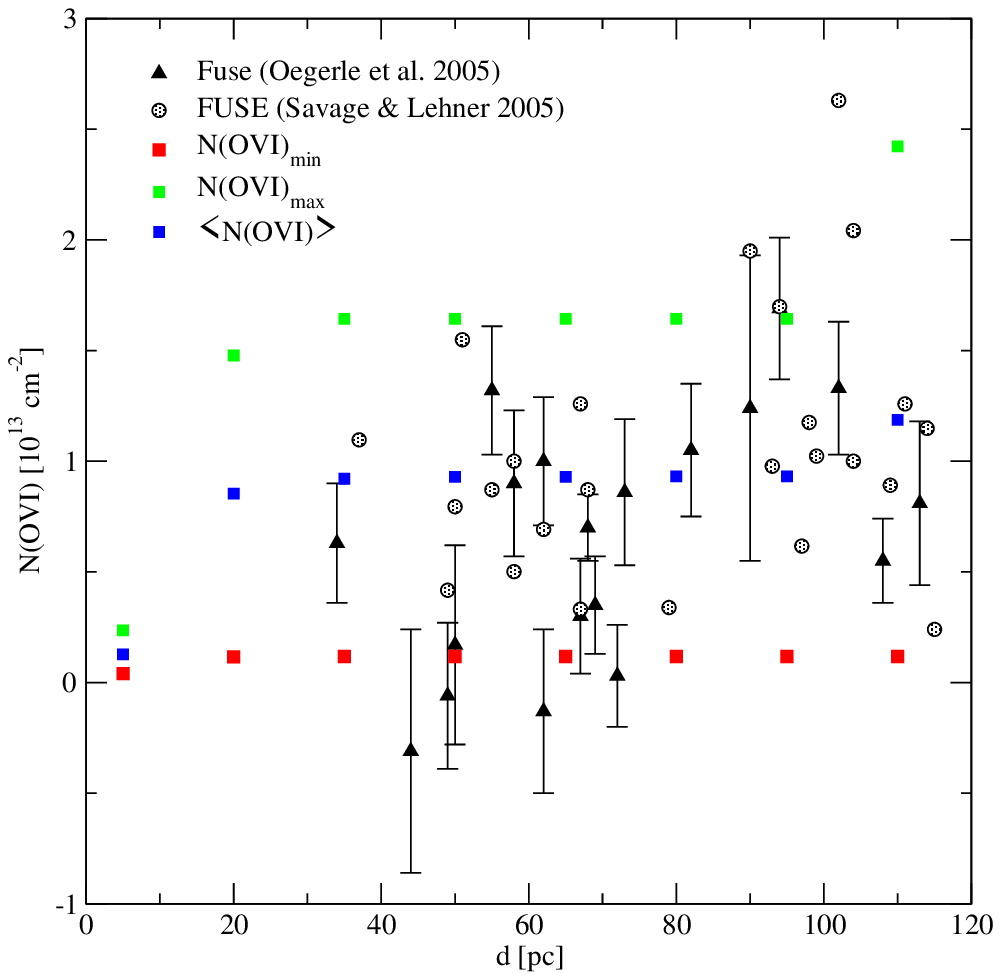}
\caption{%
Comparison between FUSE (\emph{triangles:} Oegerle et al. \cite{oe05};
\emph{circles:} Savage \& Lehner \cite{sl06} O{\sc vi} column densities
and the calculated minimum, maximum and average column densities
along the 91 lines of sight (at $t= 14.5$ Myr) pointing towards
L1 as shown in Fig.~1 (left). Negative values on the ordinate derive
from essentially zero equivalent absorption line widths, i.e.
non-detections, due to the noise in the spectrum \citep{oe05}.}
\label{FuseOegerle}
\end{figure}
In CIE, O{\sc vi} is most abundant at $T \sim 3 \times 10^5$ K and sharply peaked. We can therefore use it to calibrate the age of the LB by calculating the O{\sc vi} column density inside LB and Loop~I for an ISM plasma with solar abundances \citep{ag89}; for details see Avillez \& Breitschwerdt \cite{ab05b}. We took 91 lines of sight (LOS) extending from the Sun and crossing Loop~I (the hot pressured region
200 pc to the right of the LB) from an angle of $-45^\circ$ to
$+45^\circ$ as marked in the left panel of Fig.~\ref{fig1}.
Next we compare FUSE data (Fig.~\ref{FuseOegerle}) by Oegerle et al.
(2005; triangles \cite{oe05}) and Savage \& Lehner (2006, circles; \cite{sl06}) with
simulated minimum (red squares), maximum (green squares) and
averaged (blue squares) column densities of O{\sc vi} measurements
along the 91 LOS outlined in Fig.~\ref{fig1} (left).
It can be clearly seen that for $t \sim 14.5$
Myr the calculated N(O{\sc vi}) distribution in the LB is similar to
that observed with FUSE. We can thus constrain the age
of the LB to be about $13.5 - 14.5$ Myr, and derive then
all the relevant LB properties, such as size, density, temperature, pressure, dynamical
instabilities in the interaction shell, etc..
It has been pointed out that the low N(O{\sc vi}) values \textit{in emission} found inside the LB \citep{sh03} are a stringent test on the validity of any model.
It is therefore the subject of work in progress, and will be discussed in a forthcoming paper.

\section{Analytical Modeling}
\begin{figure}
\includegraphics[width=300pt,bb=0 58 595 807,clip=]{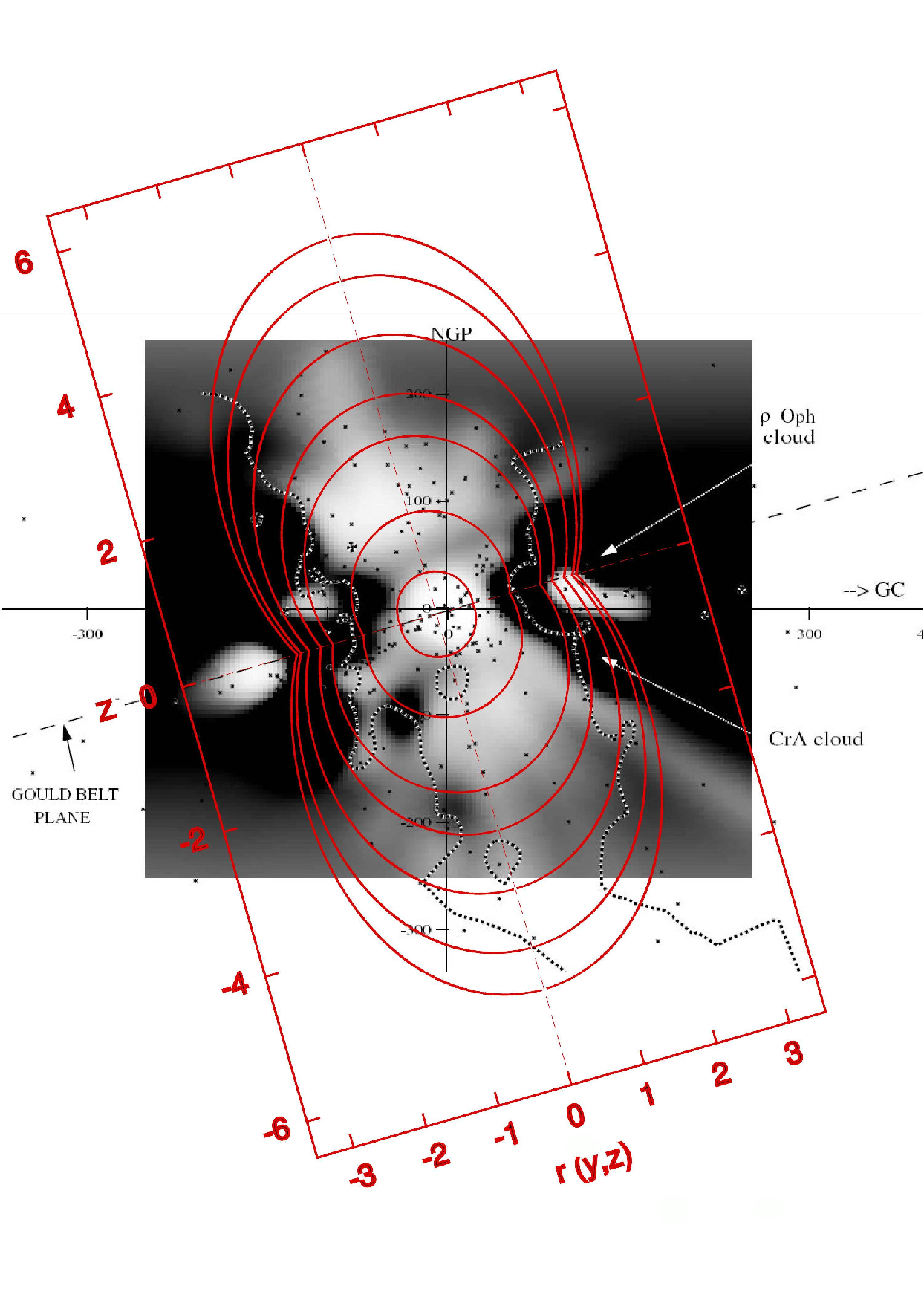}
\caption{Contours of Na{\sc i} equivalent widths corresponding to column densities of $\log $ N(H{\sc i})$=19.3$ (dashed lines, see Fig. 10 of Lallement et al. \cite{lw03}) in the meridional plane with the Kompaneets solutions overlaid. The solid line contours indicate the position of the shock front at different evolutionary stages of the bubble $y = 0.5, 1.0, 1.4, 1.55, 1.8,$ and $1.85$ with $y$ in scale height units. For a scale height of H = 70 pc and with an energy input of 19 SNe, the age of the Local Bubble can be approximated as a function of $y$ and midplane particle number density, $n_0$, by $t_{\rm{now}}(y) = 2.15 \times \left( \frac{n_0}{1\, \rm{cm}^{-3}} \right)^{0.314} y^{1.91} \, \rm{Myr}$. Obviously, the two outermost contours ($y=1.8$ and $y=1.85$) fit the size of the LB very well, corresponding to $t_{\rm{now}}=13.6$ Myr and $14.3$ Myr, respectively, taking $n_0=10\, {\rm cm}^{-3}$. Note that the inclination of $\sim 20^{\circ}$ is produced in an overlay here and is not included in the model itself.}
\label{fig:qpb}
\end{figure}

In view of the computing time consumption in 3D high resolution numerical simulations, and also in order to estimate basic properties of expanding superbubbles, it is useful to study analytical models. Since superbubbles have extensions in excess of 100 pc, the ambient density distribution, and in particular its stratification perpendicular to the midplane, cannot be neglected. Therefore self-similar models developed for the expansion into a homogeneous medium (e.g.\ \cite{wc77}, \cite{mk87}) need to be replaced by 2D models, of which Kompaneets' solution \citep{ko60} for an explosion into an exponentially stratified atmosphere seems to be the best choice. The assumptions made are: (i) strong shock, (ii) the mass is concentrated in the outer thin shell, and (iii) pressure is uniform across the bubble. Since from observations we infer that $\sim 19$ SNe have exploded during the last 14.5 Myr, we can estimate the scale height $H$ and the midplane density $\rho_0$ of the ambient distribution given by $\rho(z)=\rho_0 \cdot e^{-\vert z \vert /H}$, where $z$ is the coordinate in direction of the density gradient, which, as has been shown by Lallement et al. \cite{lw03} is tilted towards the midplane by about $\sim 20^{\circ}$. Note that the cavity is elongated symmetrically with respect to the inclined plane, which is why we use a double exponential disk. The outer shock (and shell) of such a bubble are then given in cylindrical coordinates ($r,z$) by its half-width extension parallel to the galactic plane:
\begin{equation}
r(y, z)=2H\arccos \left[ \frac{1}{2} e^{\, z/H} \left(1 - \frac{y^2}{4H^2} + e^{-z/H} \right) \right] \,,
\end{equation}
where $y=\int_{0}^{t} \sqrt{\frac{\gamma^2 - 1}{2} \frac{E_{\rm{th}}(t)}{\rho_0 \cdot V(t)}} \, dt$ is a transformed time variable, with $V(t)$ being the volume and $E_{\rm{th}}(t)$ the thermal energy input of the superbubble. The maximum half-width of a bubble is $r_{\rm{max}}(y) = 2\rm{H} \arcsin (y/2)$
and the maximum extension in $z$-direction is $z_u(y) = -2 \rm{H} \log(1-y/2)$. It is obvious that after some time RTIs will grow exponentially as the shock runs down a density gradient, leading to break-up of the shell, which according to observations of the LB (see Fig.~\ref{fig:qpb}) must have already occurred. The hot gas breaking out of the bubble will still be supersonic with respect to the ambient medium and therefore drive a shock ahead of an expanding secondary bubble or jet, so that for simplicity (and in a crude approximation) we will still use the Kompaneets' solution.

For the time-dependent energy input rate, we have taken an initial mass function (IMF) with a slope of $\Gamma = -1.1$ \citep{mj95}, and for the main-sequence life-time of massive stars we applied the more recent approximation by Fuchs et al. \cite{fu06}, relating the main sequence life time to stellar mass by $\tau_{\rm ms} = \tau_0 \, M^{-\alpha}$, with $\tau_0=1.6 \times 10^8$ yr and $\alpha =0.932$, for $2 < M < 67$ (with $M$ in units of M$_{\odot}$; for further details, see MSc thesis by Baumgartner \cite{ba07}). All SNe are assumed to explode in the center of the bubble $(r=0, z=0)$.

By comparing our model to the observed density map of \citep{lw03} we find that the contour of $y = (1.8 - 1.85) \, \rm{H}$ fits the bubble size best (see Fig.~\ref{fig:qpb}). The contours would correspond to the position of the shell if the bubble was not already broken up.
For this stage of evolution, the coordinate $z_u$ is then $(4.6 - 5.2) \, \rm{H}$ and $r_{\rm{max}}$ is $(2.2 - 2.4) \,\rm{H}$. With an extension of the bubble from the midplane to the top of $320 - 365 \,$pc, we find that the scale height is $\sim$70$ \,$pc.\\
Taking into account that the age of the LB should be roughly 14.5$ \,$Myr, we obtain a midplane number density of the undisturbed ISM of about $8 - 11 \,$ cm$^{-3}$. The break-up of the shell then started about $3.6 - 4.0 \, $Myrs after the first SN exploded, and the velocity of the polar cap of the shell at break-up was around $23 - 26$ km s$^{-1}$. The present day expansion velocity of the shell in the midplane is $\sim 5.5 \, {\rm km}\, {\rm s}^{-1}$.

\section{Summary and Conclusions}
\label{conc}
We have developed a \textit{physical model} for the LB, which makes definite predictions for the state of the local plasma, and can thus be confronted with observations. The basic features of the model are: (i) the number of SNe and their explosion times are given, (ii) the ambient medium is \textit{not} assumed to be homogeneous, but is calculated from a larger general SN driven ISM simulation, (iii) the evolution of the neighboring Loop I superbubble is calculated simultaneously, with SN explosions derived from its present stellar content, (iv) diffuse stellar background heating and radiative cooling are taken into account. We have shown that 3D high resolution numerical simulations can reproduce UV observations and the overall LB morphology for an inferred age of $13.5 - 14.5$ Myr. Moreover, the LB pressure turns out to be fairly low, i.e. $P/k_B \sim 2000 - 3000 \, {\rm K} \, {\rm cm}^{-3}$, where $k_B$ is Boltzmann's constant, in agreement with observations (for a recent discussion see \cite{je08}), resolving the earlier problem of a huge pressure imbalance between a hot LB and the Local Cloud, in which our solar system is embedded, and which has $P/k_B \sim 1300 - 3000 \, {\rm K} \, {\rm cm}^{-3}$ (e.g. \cite{fr95}). Since the last SN occurred about 0.5 Myr ago, the LB is becoming extinct and is in a state of recombination. We find at present a much smaller amount of gas at (or in excess of) $10^6$ K than any previous model, allowing for a sizeable contribution of SWCX to the soft X-ray emission. However, in a next steep we need to quantify the NEI structure and the fraction of X-rays due to delayed recombination, which is the subject of a forthcoming paper.

Comparing our LB runs to the general ISM simulations, we find it reassuring that the LB behaves quite similar to other superbubbles in the computational box, and therefore raises no questions about it being a special object, as has been hypothesized occasionally. This picture is completed by the fact that even 2D analytical solutions provide a reasonable fit to the LB morphology and correctly predict the formation of a Local Chimney.

What lies ahead is to calculate O{\sc vi} column densities \textit{in emission} and to produce soft X-ray maps in a full NEI simulation, and hope for quantitative results for the SWCX contribution from our colleagues, until the ``Local Bubble and Beyond (III)'' meeting.

\begin{theacknowledgments}
DB thanks the organizers for financial support to attend the meeting. VB acknowledges financial support from the research grant ``Computational Astrophysics'' funded by Vienna University under contract number FS 538001.
\end{theacknowledgments}


\end{document}